\documentclass[prd,superscriptaddress,amsmath,twocolumn,nofootinbib]{revtex4-2}

\pdfoutput=1
\usepackage{times}

\usepackage{graphicx}
\usepackage{xcolor}

\usepackage{array}
\usepackage{booktabs}
\usepackage{enumitem}
\usepackage{epsf,latexsym,bbm,euscript}

\usepackage{mathrsfs,mathtools,amsmath,amssymb}
\usepackage{tensor}

\makeatletter
\newcommand*{\defeq}{\mathrel{\rlap{%
			\raisebox{0.3ex}{$\m@th\cdot$}}%
		\raisebox{-0.3ex}{$\m@th\cdot$}}%
	=}
\newcommand*{\eqdef}{=\mathrel{\rlap{%
			\raisebox{0.3ex}{$\m@th\cdot$}}%
		\raisebox{-0.3ex}{$\m@th\cdot$}}%
}
\makeatother

\newcommand{\RNum}[1]{\uppercase\expandafter{\romannumeral #1\relax}}

\def\sg{\textsl{g}}

\usepackage{scalerel}
\usepackage{tikz}
\usetikzlibrary{svg.path}
\definecolor{orcidlogocol}{HTML}{A6CE39}
\tikzset{
	orcidlogo/.pic={
		\fill[orcidlogocol] svg{M256,128c0,70.7-57.3,128-128,128C57.3,256,0,198.7,0,128C0,57.3,57.3,0,128,0C198.7,0,256,57.3,256,128z};
		\fill[white] svg{M86.3,186.2H70.9V79.1h15.4v48.4V186.2z}
		svg{M108.9,79.1h41.6c39.6,0,57,28.3,57,53.6c0,27.5-21.5,53.6-56.8,53.6h-41.8V79.1z M124.3,172.4h24.5c34.9,0,42.9-26.5,42.9-39.7c0-21.5-13.7-39.7-43.7-39.7h-23.7V172.4z}
		svg{M88.7,56.8c0,5.5-4.5,10.1-10.1,10.1c-5.6,0-10.1-4.6-10.1-10.1c0-5.6,4.5-10.1,10.1-10.1C84.2,46.7,88.7,51.3,88.7,56.8z};
	}
}
\newcommand\orcidlink[1]{\href{https://orcid.org/#1}{\mbox{\scalerel*{
				\begin{tikzpicture}[yscale=-1,transform shape]
					\pic{orcidlogo};
				\end{tikzpicture}
			}{X}}}}
		
\usepackage{url,hyperref}
\hypersetup{colorlinks,linkcolor={blue!55!black},citecolor={red!50!black},urlcolor={blue!45!black},breaklinks=true}

\begin{document}

\title{Regular black holes and the first law of black hole mechanics}

\author{Sebastian Murk\orcidlink{0000-0001-7296-0420}}
\email{sebastian.murk@oist.jp}
\affiliation{Okinawa Institute of Science and Technology, 1919-1 Tancha, Onna-son, Okinawa 904-0495, Japan}

\author{Ioannis Soranidis\orcidlink{0000-0002-8652-9874}}
\email{ioannis.soranidis@hdr.mq.edu.au}
\affiliation{School of Mathematical and Physical Sciences, Macquarie University, Sydney, New South Wales 2109, Australia}

\begin{abstract}
Singularity-free regular black holes are a popular alternative to the singular mathematical black holes predicted by general relativity. Here, we derive a generic condition that spherically symmetric dynamical regular black holes must satisfy to be compatible with the first law of black hole mechanics based on an expression for the surface gravity at the outer horizon. We examine the dynamical generalizations of models typically considered in the literature and demonstrate that none of them satisfies the condition required for compatibility with the first law, suggesting that modifications are required to maintain its physical meaning. We show that the need for corrections is inherently linked to the introduction of a minimal length scale and can therefore be seen as a direct consequence of the spacetime regularization. We explicitly identify the additional work terms in the extended first law, comment on their thermodynamic interpretation, and show that the linear coefficient of the Misner--Sharp mass suffices to determine the relevant thermodynamic properties.
\end{abstract}

\maketitle

\section{Introduction} \label{sec:introduction}
In their 1973 milestone paper \cite{bch:73}, Bardeen, Carter, and Hawking introduced four laws of black hole mechanics and elucidated close analogies with the four laws of thermodynamics. This important link connecting the two fields has since proven to be a powerful tool in advancing our understanding of black holes. In particular, the physical insights revealed in the rigorous mathematical derivation of the first law in the integral and differential formalism of Ref.~\citenum{bch:73} have provided strong motivation for further investigating their thermodynamic properties \cite{b:72,b:73,b:74,h:74,h:75,p:76}.

While the existence of dark massive ultracompact objects has been established beyond any reasonable doubt, the question of whether these objects are black holes is still open \cite{v:08,b:11,h:14,f:14,cp:19,m:22b}. In the absence of a clear answer, singularity-free models of so-called regular black holes (RBHs) \cite{a:08,sz:22,lygm:23} have received much attention in recent years, as they provide a way to avoid the nontrivial causal structures inherent to the black holes predicted by general relativity (GR). Unlike the singular mathematical black holes of GR, which are bounded by globally defined and physically unobservable event horizons \cite{v:14}, RBH models are characterized by a separate inner (Cauchy) and outer (in the case of an evolving RBH spacetime quasilocal, e.g.\ apparent or trapping) horizon. Both the inner and outer horizon generally have a nonzero surface gravity. However, inner horizons with nonzero surface gravity are typically unstable under small perturbations, which gives rise to so-called mass inflation instabilities, i.e.\ exponential instabilities characterized by an exponential growth of the gravitational energy in a neighborhood of the inner horizon (as measured, for instance, by evaluating the Weyl scalar) \cite{pi:89,pi:90,o:91,ha:10,cdlpv:21,dmt:22}. Recently, a novel ``inner-extremal'' RBH model that cures the exponential instability by making the inner horizon surface gravity vanish, while maintaining the separation between the inner and outer horizon, and a nonzero surface gravity at the outer horizon, has been proposed \cite{cdlpv:22}.

It is important to note that the original 1973 paper considered stationary black holes \cite{bch:73}. While the mass loss of an evaporating black hole is usually ascribed to the emission of Hawking radiation \cite{h:74,h:75}, the backreaction of spacetime geometry is not accounted for in its derivation, which assumes that the underlying geometry is (at least asymptotically) stationary. This is a physically significant omission: if the backreaction from the Hawking flux is not ignored, variations between nearby equilibrium states are no longer accurately described by the first law.

To resolve this limitation and go beyond physically unrealistic stationary scenarios, the concept of a dynamical (and thus quasilocal) horizon has been developed \cite{h:98,ak:04,s:rev:11}. It allows us to describe the geometry of an evolving black hole spacetime and has become an indispensable mathematical tool to accurately model dynamical processes such as the formation and possible evaporation of black holes, as well as to enable generalizations of the laws of black hole mechanics and their thermodynamic interpretation \cite{h:98,kpv:21}, including that of surface gravity as a temperature parameter.

In this article, we investigate the physical consequences of the first law of black hole mechanics for dynamical models of RBHs embedded in asymptotically flat spacetimes. We restrict our considerations to the case where the entropy scales with the area of the outer horizon\footnote{Alternatives have been considered, for instance, in Refs.~\citenum{krt:10} and \citenum{pl:17}.}. Compared to the evolution of the universe, the evaporation of black holes is considered to be a thermodynamically slow process \cite{h:16}. Therefore, if the first law is true, the behavior of dynamically evolving black holes should match that prescribed by the first law in the quasistatic limit. Based on this assertion, we derive a generic condition that RBHs must satisfy to be compatible with the first law by considering the surface gravity at the outer horizon. We explicitly test the dynamical generalizations of popular models such as those proposed by Bardeen \cite{b:68} (Subsec.~\ref{subsec:B.RBH}), Dymnikova \cite{d:92} (Subsec.~\ref{subsec:D.RBH}), Hayward \cite{h:06} (Subsec.~\ref{subsec:H.RBH}), the model considered in Ref.~\citenum{cdddos:23} (Subsec.~\ref{subsec:sSc.RBH}) which produces the strongest possible corrections to the Schwarzschild geometry while still being compatible with its asymptotic behavior, and the aforementioned inner-extremal RBH model \cite{cdlpv:22} (Sec.~\ref{sec:test.deg}). Our analysis shows that none of these models is compatible with the conventional form of the first law of black hole mechanics.

The remainder of this article is organized as follows: In Sec.~\ref{sec:math.prereq}, we introduce mathematical concepts used in the construction of RBHs (Subsec.~\ref{subsec:trapped_reg}) and review the first law of black hole mechanics and its relation to surface gravity (Subsec.~\ref{subsec:SG.FL}). In Sec.~\ref{sec:comp.cond}, we derive a generic condition that dynamical black holes must satisfy to be compatible with the first law of black hole mechanics. Based on this compatibility condition, we test the dynamical generalizations of commonly considered RBH models, and find that none of them satisfies the required relation (Sec.~\ref{sec:test.nondeg} and Sec.~\ref{sec:test.deg}), suggesting that either these models do not conform to the first law or modifications of the first law are required to maintain its essence. In Sec.~\ref{sec:Page.law}, we briefly outline the consequences of this result in the context of the so-called Page evaporation law. Lastly, we discuss the implications of our findings more generally and comment on possible directions for future research related to nonsingular black hole spacetimes (Sec.~\ref{sec:concl}). Throughout this article, we use the metric signature $(-,+,+,+)$ and work in dimensionless units such that $c=G=\hbar=k_B=1$.

\section{Mathematical prerequisites} \label{sec:math.prereq}

\subsection{Trapped regions and regular black holes} \label{subsec:trapped_reg}
A general spherically symmetric metric in advanced null coordinates $(v,r)$ is described by the line element
\begin{align}
	ds^2=-e^{2h(v,r)}f(v,r)dv^2+2e^{h(v,r)}dvdr+r^2d\Omega^2 ,
	\label{eq:metric}
\end{align}
where $d\Omega^2 = d\theta^2 + \sin^2 \theta \, d\phi^2$ denotes the line element of the 2-sphere. Since our argumentation in Subsec.~\ref{subsec:SG.FL} and Sec.~\ref{sec:comp.cond} is based on the analysis of surface gravity at the outer horizon [i.e.\ at $r=r_+(v)$] and it is always possible to write the function $h(v,r)$ as a series with respect to the coordinate distance $r-r_+(v)$ from the outer horizon (see Ref.~\citenum{mmt:rev:22} for details),
\begin{align}
	h(v,r) &= \sum\limits_{i=1}^\infty \chi_i(v) \big(r-r_+(v)\big)^i ,
\end{align}
we can assume $h(v,r)=0$ without loss of generality in what follows\footnote{Note also that $h=0$ for all RBH models typically considered in the literature, including all of those examined explicitly in Secs.~\ref{sec:test.nondeg} and \ref{sec:test.deg}.}. Generic dynamical models of RBHs are then described by the metric function
\begin{align}
	f(v,r) \defeq g(v,r)\big(r-r_{-}(v)\big)^a\big(r-r_{+}(v)\big)^b ,
	\label{eq:fvr.gen}
\end{align}
where $r_{-}(v)$ and $r_{+}(v)$ denote the inner and outer horizon, respectively, and $a,b \in \mathbb{N}_{>0}=\lbrace1,2,...\rbrace$ are positive integers labeling their degeneracy. In spherical symmetry, nonsingular black hole metrics possess an inner horizon due to the fact that the outer horizon (which is located close to the classical gravitational radius) cannot cross the center $r=0$ without creating a curvature singularity \cite{f:16}. The inner horizon is generated by the additional hair, i.e.\ the minimal length scale $l$ that is introduced to regularize the spacetime [cf.\ Eqs.~\eqref{eq:H.RBH.rm}, \eqref{eq:B.RBH.rm}, \eqref{eq:D.RBH.rm}, and \eqref{eq:sSc.RBH.rm}] and possibly other parameters such as charge [cf.\ Eqs.~\eqref{eq:RNBH.dyn.rm} and \eqref{eq:HF.RBH.rm(v,l)}]. Constraints for the \textit{a priori} undetermined function $g(v,r)$ are discussed in what follows. 

The expansions of ingoing and outgoing radial null geodesic congruences are given by
\begin{align}
	\theta_{-}=-\frac{2}{r} \, , \quad \theta_{+}=\frac{f(v,r)}{r} \, ,
	\label{eq:rad.null.geod.exp}
\end{align}
respectively. The existence of a trapped spacetime region is contingent on the signature of their product $\theta_{-} \theta_{+} \lessgtr 0$. We follow the widely used convention proposed in Ref.~\citenum{h:94}, according to which the presence of a trapped region bounded by the outer horizon $r_+$ is signified by $\theta_{-} \theta_{+} > 0$ (i.e.\ the future-directed expansions of both ingoing and outgoing null geodesics are negative), and no trapped region is present for $\theta_{-} \theta_{+} < 0$. Since $\theta_-$ is always negative, this implies that $f<0$ inside of the trapped region $r \in (r_-,r_+)$, and $f>0$ outside of the trapped region $r > r_+$, and thus $g>0$ and $b$ odd (otherwise $g$ would have to be negative inside, but positive outside of the trapped region). The inner and outer horizon are identified as the roots of the equation $f=0$ \cite{fefhm:17,bhl:18}. At the ``disappearance point'' of the trapped region $v=v_d$, they coalesce, i.e.\ $r_-(v_d) \equiv r_+(v_d)$. From Eqs.~\eqref{eq:fvr.gen} and \eqref{eq:rad.null.geod.exp}, it then follows that 
\begin{align}
	\theta_{-} \theta_{+} \big\vert_{v=v_d} = -\frac{2}{r^2}g(v_d,r)\big(r-r_{+}(v_d)\big)^{a+b} \leq 0 \quad \forall r \; ,
\end{align}
which implies that the sum $a+b$ must be even, and thus $a$ odd. 

We note that the formation of a trapped spacetime region in finite time according to the clock of a distant observer inevitably requires a violation of the null energy condition (NEC) near the outer horizon \cite{he:book:73,fn:book:98,bhl:18,bmmt:19}, which posits that $\tensor{T}{_\mu_\nu} \ell^\mu \ell^\nu \geqslant 0$, i.e.\ the contraction of the energy-momentum tensor with any future-directed null vector $\ell^\mu$ is non-negative. Similarly, violating the NEC is a prerequisite for the emission of Hawking radiation. While quantum effects are necessary, it is worth noting that Hawking radiation is a purely kinematical phenomenon \cite{v:03}, and neither the Einstein equations nor the Bekenstein entropy relation \cite{b:72,b:73,b:74} are required for its derivation.

\subsection{Surface gravity and the first law of black hole mechanics} \label{subsec:SG.FL}
The first law of black hole mechanics derived in Ref.~\citenum{bch:73} has been proven to hold in any theory of gravity arising from a diffeomorphism-invariant Lagrangian \cite{w:93,iw:94}. Assuming $\delta J = \delta Q = 0$, it can be stated mathematically as
\begin{align}
	\delta M &= \frac{\kappa}{8 \pi} \delta A ,
	\label{eq:first_law}
\end{align}
where $M$, $\kappa$, and $A$ denote the black hole's gravitational energy, surface gravity, and horizon area, respectively. The notion of gravitational energy within a sphere of radius $r$ is captured by the so-called Misner--Sharp (MS) mass \cite{ms:64} $M \defeq C/2$ [see Eqs.~\eqref{eq:f.def.MSmass} and \eqref{eq:MSmass}]\footnote{While the MS mass is technically $C/2$ by virtue of its definition in Eq.~\eqref{eq:f.def.MSmass}, we often take the liberty to refer to $C$ itself simply as the MS mass.}. In spherically symmetric solutions of the Einstein equations, such as the static Schwarzschild\footnote{Note that in the Schwarzschild solution, $M=\mathrm{const.}$ corresponds to the Arnowitt--Deser--Misner (ADM) mass, and the object's Schwarzschild radius $r_\sg$ corresponds to the outer horizon, $r_\sg \equiv r_+ = 2M$.} or the nonstatic Vaidya metric, $C(v,r_+) = 2M(v) = r_+(v)$. Using $A = 4 \pi r_+^2$ for the horizon area, this leads to the famous expression for the surface gravity at the outer horizon:
\begin{align}
	\frac{\delta M}{\delta r_+} = \frac{\kappa}{8 \pi} \frac{\delta A}{\delta r_+} \; \Rightarrow \; \kappa = \frac{1}{2 r_+} .
	\label{eq:firstlaw_sg}
\end{align}
It is important to note that the surface gravity $\kappa$ is unambiguously defined only in stationary spacetimes, where it is related to the black hole's Hawking temperature $T_H$ via
$\kappa = 2 \pi T_H$.
Nonetheless, even in generic dynamical black hole spacetimes the first law and its associated expression for the surface gravity are expected to approach Eqs.~\eqref{eq:first_law} and \eqref{eq:firstlaw_sg} in the quasistatic limit due to the timescale of the evaporation process. We show here that this is not the case for the RBH models typically considered in the literature.

Generalizations of surface gravity to dynamical spacetimes \cite{ny:08,clv:13} are generally related to either the affine peeling surface gravity \cite{blsv:11} or the so-called Kodama surface gravity \cite{k:80,av:10,kpv:21}. Since the peeling surface gravity is ill-defined for a transient object that forms in finite time of a distant observer \cite{mt:21a,mt:21b,mmt:21}, and there are strong arguments that Kodama surface gravity is the critical quantity with respect to Hawking radiation \cite{kpv:21,mmt:21}, we focus on this generalization of surface gravity in what follows. 

The main difficulty in the generalization of surface gravity to evolving black hole spacetimes is that, unlike their stationary counterparts, they are not guaranteed to admit a timelike Killing vector field that generates the null hypersurface (known as Killing horizon) needed to define surface gravity. However, a dynamical notion of surface gravity can be defined at the (quasilocal) outer horizon using the Kodama vector field \cite{k:80,av:10}, which is well-defined even in nonstationary spherically symmetric spacetimes, and thus in some sense supersedes the concept of a Killing vector field.

At the outer horizon, the Kodama surface gravity $\kappa_K$ is defined via
\begin{align}
	\kappa_K K_\nu \eqdef \frac{1}{2} K^\mu \big( \nabla_\mu K_\nu - \nabla_\nu K_\mu \big) ,
	\label{eq:Kod.SG.def}
\end{align}
where $K^\mu$ denotes the contravariant Kodama vector \cite{k:80,av:10}. The Kodama vector field is conserved,
\begin{align}
	\nabla_\mu K^\mu = 0 ,
\end{align}
and generates a conserved current
\begin{align}
	\nabla_\mu J^\mu = 0 \, , \quad J^\mu \defeq \tensor{G}{^\mu^\nu} K_\nu \, ,
\end{align}
for any symmetric rank-2 tensor $\tensor{G}{_\mu_\nu}=\tensor{G}{_\nu_\mu}$ that is invariant under the spherical symmetries of the spacetime. If $\tensor{G}{_\mu_\nu}$ is the Einstein tensor, then the current's Noether charge is the MS mass.

\section{Derivation of the compatibility condition} \label{sec:comp.cond}
For the metric specified in Eq.~\eqref{eq:metric} (recall that, as established in Subsec.~\ref{subsec:trapped_reg}, we can use $h=0$ without loss of generality), $K^\mu = (1,0,0,0)$, and its only nonzero covariant component at the outer horizon is $K_r = 1$. Hence the Kodama surface gravity [cf.\ Eq.~\eqref{eq:Kod.SG.def}] at the outer horizon is given by
\begin{align}
	\kappa_K \big\vert_{r=r_+} &= \frac{1}{2} \partial_r f(v,r)  \big\vert_{r=r_+} \label{eq:KodSG} \\
	& \stackrel{\eqref{eq:fvr.gen}}{=}\lim_{r\rightarrow r_{+}} \frac{(r-r_{+})^{-1+b} b g(v,r) (r - r_-)^a}{2} , \label{eq:KodSG.lim}
\end{align}
which implies that a nonzero Kodama surface gravity at the outer horizon can be achieved only if the outer horizon is nondegenerate, i.e.\ $b=1$. We thus focus on this scenario in what follows.

Assuming that $f(v,r)$ is decomposable as a rational function of the radial coordinate $r$, i.e.\
\begin{align}
	f(v,r) &= \frac{\mathcal{P}_n(r)}{\tilde{\mathcal{P}}_n(r)} ,
	\label{eq:f(v,r).poly.decomp}
\end{align}
where $\mathcal{P}_n$ and $\tilde{\mathcal{P}}_n$ are polynomials (whose coefficients depend on $v$ in generic dynamical spacetimes) of the same degree $n\geqslant3$ in $r$ as motivated in Ref.~\citenum{f:16}, we can write
\begin{align}
	g(v,r) &= \frac{\sum\limits_{z=0}^{m} \lambda_z r^z}{\sum\limits_{i=0}^{n} c_i r^i} ,
	\label{eq:g(v,r).decomp}
\end{align}
where the coefficients $\lambda_z \equiv \lambda_z(r_-,r_+)$ and $c_i \equiv c_i(r_-,r_+)$ depend explicitly only on $r_-(v)$ and $r_+(v)$ (and thus implicitly on $v$) as they are the only relevant length scales, and $n - m = a + 1$ and $\lambda_m/c_n = 1$ are required to recover the Vaidya form of the metric in the asymptotic limit $r \to \infty$. These considerations will prove useful in our analysis of the inner-extremal RBH model in Sec.~\ref{sec:test.deg}. We also note that --- complemented by the assumptions of regularity of the spacetime at the origin $r=0$ and a proper Schwarzschild/Vaidya form of the metric in the asymptotic limit --- the polynomial decomposition of the metric function according to Eq.~\eqref{eq:f(v,r).poly.decomp} immediately leads to a class of metric families of the form (see Ref.~\citenum{f:16} for a detailed derivation)
\begin{align}
	f(v,r) &= 1 - \frac{r_\sg(v) r^2}{r_\sg(v) l(v)^2 + c_1(v) r + c_2(v) r^2 + r^3}
\end{align}
for the case $n=3$, where $l(v)$ denotes the minimal length scale, $r_\sg(v)=2M(v)$, and the case $c_1(v)=c_2(v)=0$ corresponds to the dynamical Hayward metric of Eq.~\eqref{eq:H.RBH.f(v,r)} considered in Subsec.~\ref{subsec:H.RBH}.

The metric function $f$ [cf.\ Eq.~\eqref{eq:fvr.gen}] is usually defined in terms of the MS mass via
\begin{align}
	f(v,r) \defeq \partial_\mu r \partial^\mu r = 1 - \frac{C(v,r)}{r} ,
	\label{eq:f.def.MSmass}
\end{align}
with the MS mass given by
\begin{align}
	C(v,r) = r_+(v) + \sum\limits_{i=1}^\infty w_i(v) \big(r-r_+(v)\big)^i .
	\label{eq:MSmass}
\end{align}
By means of Eq.~\eqref{eq:f.def.MSmass}, the Kodama surface gravity of Eq.~\eqref{eq:KodSG} can then be expressed directly in terms of the MS mass as
\begin{align}
	\kappa_K \big\vert_{r=r_+} \hspace{-1mm} &= \frac{1}{2 r^2} \big[ C(v,r) - r \partial_r C(v,r) \big] \big\vert_{r=r_+} \hspace{-1.25mm} \stackrel{\eqref{eq:MSmass}}{=} \frac{1-w_1}{2r_+} ,
	\label{eq:KodamaSGv2}
\end{align}
where the rightmost expression is obtained by explicitly substituting the MS mass expansion of Eq.~\eqref{eq:MSmass}. Evaluation at the outer horizon yields
\begin{align}
	\delta M \big\vert_{r=r_+} = \left. \frac{\delta C}{2} \right|_{r=r_+} = \frac{1}{2} \left(1-w_1\right) \delta r_+ .
	\label{eq:MS_outer_horizon}
\end{align}
Substituting the expression for the surface gravity on the RHS of Eq.~\eqref{eq:KodamaSGv2} into the RHS of Eq.~\eqref{eq:first_law}, we obtain 
\begin{align}
	\delta M &= \frac{1-w_1}{16 \pi r_+} \delta A .
	\label{eq:first_law_expl}
\end{align} 
Subsequent substitution of Eq.~\eqref{eq:MS_outer_horizon} into Eq.~\eqref{eq:first_law_expl} leads to
\begin{align}
	\delta \left(\frac{r_{+}}{2}\right)=\frac{1-w_1}{16 \pi r_{+}} \delta A+\frac{w_1}{2} \delta r_{+} .
\end{align}
Using the relation $\delta A = 8 \pi r_+ \delta r_+ = \tfrac{2}{r_+} \delta V$ between area $A=4\pi r_+^2$ and volume $V=\tfrac{4}{3}\pi r_+^3$, we obtain
\begin{align}
	\delta \left(\frac{r_{+}}{2}\right)=\frac{1-w_1}{16 \pi r_{+}} \delta A + \frac{w_1}{8 \pi r_{+}^2} \delta V ,
	\label{eq:first_law_extended}
\end{align}
which is an extension of the standard form of the first law of black hole mechanics that includes an additional work term. The notion of internal/thermal energy is captured by the expression $r_+/2$, which corresponds to the MS mass $C/2$ evaluated at the outer horizon. However, this internal energy is not necessarily the same as the ADM mass when matter fields are present \cite{mz:14}, as is the case for dynamically evolving black holes. Comparison with the traditional form of the first law of mechanics $\delta M = \tfrac{\kappa}{8 \pi} \delta A - p \delta V$ \cite{lm:22} identifies the pressure $p$ in terms of $w_1$, i.e.\
\begin{align}
	p = - \frac{w_1}{8 \pi r_+^2} .
	\label{eq:pressure_term}
\end{align}	

A few brief comments are in order: first, it is important to note that higher-order terms $i>1$ in the MS mass expansion have not been neglected in this derivation, but rather, they simply do not enter the expression for the Kodama surface gravity of Eq.~\eqref{eq:KodamaSGv2}. Second, the conventional form of the first law of black hole mechanics [Eq.~\eqref{eq:first_law}] and its associated expression for the surface gravity $\kappa=1/(2r_+)$ [cf.\ Eq.~\eqref{eq:firstlaw_sg}] are attainable only if $w_1=0$, as can be seen from Eqs.~\eqref{eq:KodamaSGv2} and \eqref{eq:first_law_extended}. Therefore, 
\begin{align}
	\boxed{w_1 \big\vert_{r=r_+} \hspace{-0.75mm} = 0} 
	\label{eq:w1=0}
\end{align}
is a necessary condition to be compatible with the first law, i.e.\ the linear coefficient in the MS mass expansion $w_1$ [cf.\ Eq.~\eqref{eq:MSmass}] must vanish at the outer horizon. Physically, this implies that the metric approximates the Vaidya solution near the outer horizon. Third, the expressions derived in Eqs.~\eqref{eq:KodamaSGv2}--\eqref{eq:w1=0} apply generically to black holes described by a metric function of the form of Eq.~\eqref{eq:f.def.MSmass}. For RBHs described by a metric function of the form of Eq.~\eqref{eq:fvr.gen}, performing a series expansion of Eq.~\eqref{eq:MSmass} about the outer horizon yields
	\begin{align}
		w_1 \big\vert_{r=r_+} \hspace{-0.75mm}= 1 - g(v,r_+) r_+ (r_+ - r_-)^a .
		\label{eq:w1}
\end{align}
If the inner horizon is nondegenerate ($a=1$), as is typically the case for the RBHs most commonly considered in the literature (e.g.\ those of Refs.~\citenum{b:68,d:92,h:06,cdddos:23} examined in Sec.~\ref{sec:test.nondeg}), the compatibility condition prescribed by Eq.~\eqref{eq:w1=0} can be evaluated straighforwardly from the expression derived in Eq.~\eqref{eq:w1} by considering the series expansion of the MS mass about the outer horizon using Eq.~\eqref{eq:f.def.MSmass} since the metric function $f$ (and thus by extension $g$) is known explicitly. The case of a degenerate inner horizon ($a>1$) is treated in Sec.~\ref{sec:test.deg} on the basis of the inner-extremal RBH model proposed in Ref.~\citenum{cdlpv:22}, where $\sum_z \lambda_z r^z = 1$ [cf.\ Eq.~\eqref{eq:g(v,r).decomp}]. In this case, the equation $f=0$ has two positive real-valued solutions, and the smaller one which corresponds to the inner horizon is degenerate. If the outer horizon $r=r_+$ is a single root, then the inner horizon root $r=r_-$ has to be at least cubic in order for the inner horizon surface gravity to vanish, which plays a crucial role in preventing mass inflation instabilities and ensuring that the backreaction of perturbations vanishes asymptotically.

In the next two sections, we investigate whether the consistency condition Eq.~\eqref{eq:w1=0} that is required to be compatible with the first law of black hole mechanics is satisfied by the RBH models typically considered in the literature. Since, as argued in Sec.~\ref{sec:introduction}, dynamical models are necessary to accurately model the evolution of evaporating black holes, our analysis in Secs.~\ref{sec:math.prereq} and \ref{sec:comp.cond} was designed to accommodate generic dynamical RBH models, and we consider the dynamical generalizations of popular models rather than their static counterparts in what follows. To be more precise, static RBH metrics belongs to a different class of black hole solutions whose effective energy-momentum-tensor component scaling behavior $\tensor{\tau}{_\mu_\nu} \sim f^k$ close to the outer horizon is characterized by $k=1$ as opposed to $k=0$ (see chapter 2 in Ref.~\citenum{mmt:rev:22} for a detailed exposition of the two classes of admissible solutions in spherical symmetry, or Table \RNum{1} in Ref.~\citenum{m:22a} for a succinct overview). While the unique $k=1$ solution describes black holes at the instant of their formation, they are described by a $k=0$ solution for the remainder of their entire evolution or until their disappearance (e.g.\ through complete evaporation) \cite{mt:21a}. The relation between the Kodama surface gravity at the outer horizon and the linear coefficient $w_1$ of the MS mass that we derived in Eq.~\eqref{eq:KodamaSGv2} is valid for $k=0$ solutions, which includes all dynamical models. However, it no longer applies (at least not in general) to static metrics such as the Reissner--Nordström solution briefly considered in Subsec.~\ref{subsec:HF.RBH} [cf.\ Eq.~\eqref{eq:RNbh.SG}].

While we strive to be as generic as possible in our analysis, recall that a nondegenerate outer horizon ($b=1$) is necessary to allow for a nonzero Kodama surface gravity at the outer horizon [cf.\ Eq.~\eqref{eq:KodSG.lim}]. However, in Sec.~\ref{sec:test.deg} we do allow for the possibility that the inner horizon is degenerate ($a>1$), as such models have been shown to exhibit interesting physical properties making it possible to evade the mass inflation problem that typically plagues RBHs (as detailed in Ref.~\citenum{cdlpv:22}).

\section{Testing regular black holes part \RNum{1}: models with a nondegenerate inner horizon} \label{sec:test.nondeg}
In this section, we proceed by examining RBH models with a nondegenerate inner horizon ($a=1$), including the proposals of Bardeen \cite{b:68} (Subsec.~\ref{subsec:B.RBH}), Dymnikova \cite{d:92} (Subsec.~\ref{subsec:D.RBH}), and Hayward \cite{h:06} (Subsec.~\ref{subsec:H.RBH}). We write explicit dependencies on $v$ only once when specifying the metric function $f(v,r)$ for the generalized dynamical model and omit them thereafter for the sake of readability. However, we later reinstate the explicit dependencies in Subsec.~\ref{subsec:HF.RBH} for clarity.

\subsection{A simple nonsingular ``dynamical'' spacetime} \label{subsec:sRBH}
Arguably the simplest nonsingular dynamical spacetime one can construct based on Eq.~\eqref{eq:fvr.gen} is given by the choice $a=b=1$ and $\sum_z \lambda_z r^z = 1$ [cf.\ Eq.~\eqref{eq:g(v,r).decomp}], such that Eq.~\eqref{eq:fvr.gen} can be written as
\begin{align}
	f(v,r) &= g(v,r)\big(r-r_-(v)\big)\big(r-r_+(v)\big) ,
	\label{app:eq:simple_model_f(v,r)}
\end{align}
and, by virtue of Eq.~\eqref{eq:f(v,r).poly.decomp},
\begin{align}
	g(v,r) &= \frac{1}{c_0 + c_1 r + c_2 r^2} ,
	\label{app:eq:simple_model_g(v,r)}
\end{align}
where $c_2 = 1$ is necessary to recover the Vaidya form of the metric in the asymptotic limit. Regularity at the center $r=0$ requires that
\begin{align}
	c_0 &= r_- r_+ , \; \; c_1 = - r_- - r_+ , \label{eq:sRBH.c0c1}
\end{align}
as can be verified by evaluating the Ricci ($\tensor{\sg}{_\mu_\nu} \tensor{R}{^\mu^\nu}$) or the Kretschmann ($\tensor{R}{_\mu_\nu_\rho_\sigma} \tensor{R}{^\mu^\nu^\rho^\sigma}$) scalar. However, substitution of Eqs.~\eqref{app:eq:simple_model_g(v,r)} and \eqref{eq:sRBH.c0c1} into Eq.~\eqref{app:eq:simple_model_f(v,r)} reveals that these coefficients lead to a trivial metric function
\begin{align}
	f(v,r) &= \frac{(r-r_-)(r-r_+)}{r_- r_+ - (r_- + r_+) r + r^2} = 1 . \label{eq:sRBH.f(v,r).triv}
\end{align}
In other words, we have rediscovered flat Minkowski spacetime.

\subsection{Hayward model} \label{subsec:H.RBH}
A nontrivial minimal RBH model that reduces to a de Sitter spacetime in the limit $r \to 0$ and a Schwarzschild/Vaidya spacetime in the limit $r \to \infty$ was proposed by Hayward in Ref.~\citenum{h:06}. In this model, the metric is specified by the function
\begin{align}
	f(r) &= 1 - \frac{r_\sg r^2}{r^3 + r_\sg l^2} ,
	\label{app:eq:f(r)_Hayw}
\end{align}
where $l \geqslant 0$ represents a minimal length scale (which can be interpreted as an additional hair of the black hole) akin to a Planckian cutoff, $l=0$ corresponds to the Schwarzschild solution, and $r_\sg=2M$ denotes the black hole's Schwarzschild radius. The horizons are identified through the equation $f(r)=0$ \cite{fefhm:17,bhl:18}, which can be solved using Cardano's formula \cite{w:37} and admits three real solutions, namely
\begin{align}
	r_0 &= -l + \frac{l^2}{2r_\sg} + \mathcal{O}\big(l^3\big) < 0 , \label{eq:H.RBH.r0} \\
	r_- &= l + \frac{l^2}{2r_\sg} + \mathcal{O}\big(l^3\big) , \label{eq:H.RBH.rm} \\
	r_+ &= r_\sg - \frac{l^2}{r_\sg} + \mathcal{O}\big(l^4\big) . \label{eq:H.RBH.rp} 
\end{align}
Note that this is a necessary requirement in order for the metric function of Eq.~\eqref{app:eq:f(r)_Hayw} to describe a RBH, as one real and two complex solutions would imply that no inner horizon is present \cite{bhl:18}. 

We can generalize this model to describe a dynamical RBH spacetime by allowing for an explicit dependence of $r_\sg$ and $l$ on $v$, i.e.\
\begin{align}
	f(v,r) &= 1 - \frac{r_\sg(v) r^2}{r^3 + r_\sg(v) l(v)^2} .
	\label{eq:H.RBH.f(v,r)}
\end{align}
Using the roots of $f(v,r)=0$, this can be rewritten as
\begin{align}
	f(v,r) &= \frac{r-r_0}{r^3 + r_\sg l^2} (r-r_-)(r-r_+) .
\end{align}
By comparison with Eq.~\eqref{eq:fvr.gen}, we identify
\begin{align}
	g(v,r) &= \frac{r-r_0}{r^3 + r_\sg l^2} > 0 ,
\end{align}
which is positive due to $r_0<0$. The MS mass $C(v,r)$ [cf.\ Eq.~\eqref{eq:MSmass}] for the generalized dynamical Hayward model is specified by Eq.~\eqref{eq:H.RBH.f(v,r)} through the relation given in Eq.~\eqref{eq:f.def.MSmass}. By considering its expansion about the outer horizon in the regime $l \ll 1$, we obtain the linear coefficient
\begin{align}
	w_1 \big\vert_{r=r_+} \hspace{-1mm} & \stackrel{\eqref{eq:H.RBH.rp}}{=} \hspace{0.5mm} \frac{3 l^2}{r_\sg^2} + \mathcal{O}\big(l^4\big) \geqslant 0 . \label{eq:H.RBH.w1}
\end{align}
This expression is zero only if $l=0$, but then $r_-=0$ [cf.\ Eq.~\eqref{eq:H.RBH.rm}], indicating that there is no inner horizon. Consequently, the dynamical Hayward model specified by Eq.~\eqref{eq:H.RBH.f(v,r)} cannot satisfy the compatibility condition Eq.~\eqref{eq:w1=0} while $l \neq 0$.

\subsection{Bardeen model} \label{subsec:B.RBH}
The first nonsingular black hole spacetime was proposed by Bardeen in 1968 \cite{b:68}. It is defined by the metric function
\begin{align}
	f(r) = 1 - \frac{r_\sg r^2}{\big(r^2+l^2\big)^{3/2}} ,
\end{align}
where $r_\sg$ and $l$ have the same physical meaning as in Subsec.~\ref{subsec:H.RBH}. Once again, we consider its dynamical generalization
\begin{align}
	f(v,r) = 1 - \frac{r_\sg(v) r^2}{\big(r^2+l(v)^2\big)^{3/2}} .
\end{align}
Writing the inner and outer horizon in terms of $r_\sg$ and $l$, we find
\begin{align}
	r_- &= \frac{l^{3/2}}{\sqrt{r_\sg}} + \mathcal{O}\big(l^{5/2}\big) , \label{eq:B.RBH.rm} \\
	r_+ &= r_\sg - \frac{3 l^2}{2 r_\sg} + \mathcal{O}\big(l^4\big) . \label{eq:B.RBH.rp}
\end{align}
Following the same steps as in Subsec.~\ref{subsec:H.RBH}, we obtain
\begin{align}
	w_1 \big\vert_{r=r_+} \hspace{-1mm} \stackrel{\eqref{eq:B.RBH.rp}}{=} \hspace{0.5mm} \frac{3 l^2}{r_\sg^2} + \mathcal{O}\big(l^4\big) \geqslant 0 ,
\end{align}
which coincides with the linear MS mass coefficient of the dynamical Hayward model [cf.\ Eq.~\eqref{eq:H.RBH.w1}] at leading order, although higher-order contributions will differ once sufficiently high orders $\mathcal{O}(l^x)$ in Eqs.~\eqref{eq:H.RBH.rp} and \eqref{eq:B.RBH.rp} are taken into account. Similar to the dynamical Hayward model discussed in the previous subsection, this expression is only zero if $l=0$ and thus $r_-=0$ [cf.\ Eq.~\eqref{eq:B.RBH.rm}].

\subsection{Dymnikova model} \label{subsec:D.RBH}
Another well-known RBH model was proposed by Dymnikova \cite{d:92}. It is specified by the metric function
\begin{align}
	f(r) = 1 - \frac{r_\sg \big(1-e^{-r^3/r_\star^3}\big)}{r} ,
\end{align}
where $r_\star^3 \defeq l^2 r_\sg$, $r_\sg$ denotes the Schwarzschild radius, and $l$ a minimal length parameter. A generalized dynamical version is given by
\begin{align}
	f(v,r) = 1 - \frac{r_\sg(v) \big(1-e^{-r^3/r_\star(v)^3}\big)}{r} ,
\end{align}
and the inner and outer horizon can be expressed in terms of the parameters $l$ and $r_\sg$ as
\begin{align}
	r_- &= l \Big(1+\mathcal{O}\big(e^{-r_\sg^2/l^2}\big)\Big) , \label{eq:D.RBH.rm} \\
	r_+ &= r_\sg \Big(1+\mathcal{O}\big(e^{-r_\sg^2/l^2}\big)\Big) . \label{eq:D.RBH.rp} 
\end{align}
From the expansion of the MS mass about the outer horizon, we obtain its linear coefficient
\begin{align}
	w_1 \big\vert_{r=r_+} \hspace{-1mm} \stackrel{\eqref{eq:D.RBH.rp}}{=} \hspace{0.5mm} \frac{3 r_\sg^2}{l^2} e^{-\tfrac{r_\sg^2}{l^2}} +\mathcal{O}\big(e^{-r_\sg^2/l^2}\big) \geqslant 0 ,
\end{align}
which is strictly greater than zero provided that $l \neq 0$, which would once again imply that the inner horizon is absent, i.e.\ $r_-=0$ [cf.~Eq.~\eqref{eq:D.RBH.rm}].

\subsection{RBH model with the strongest Schwarzschild corrections} \label{subsec:sSc.RBH}
The RBH considered in Ref.~\citenum{cdddos:23} exhibits the strongest possible corrections to the Schwarzschild geometry while still being compatible with its asymptotics. It is described by the metric function
\begin{align}
	f(r) &= 1 - \frac{r_\sg r^2}{(r+l)^3} ,
\end{align}
and is of particular interest as observational data of the S2 star orbiting Sgr A$^\star$ can be used to test its geometry and derive upper bounds for the new length scale $l$. We once again generalize this metric by allowing for an explicit time dependence, i.e.\
\begin{align}
	f(v,r) &= 1 - \frac{r_\sg(v) r^2}{\big(r+l(v)\big)^3} .
\end{align}
Using the same procedure as in the previous subsections, we find that the inner and outer horizon are given by
\begin{align}
	r_- &= \frac{l^{3/2}}{\sqrt{r_\sg}} + \frac{3l^2}{2r_\sg} + \mathcal{O}\big(l^{5/2}\big) , \label{eq:sSc.RBH.rm} \\
	r_+ &= r_\sg - 3l - \frac{3l^2}{r_\sg} + \mathcal{O}\big(l^3\big) . \label{eq:sSc.RBH.rp} 
\end{align}
For the linear coefficient of the MS mass at the outer horizon, we find
\begin{align}
	w_1 \big\vert_{r=r_+} \hspace{-1mm} \stackrel{\eqref{eq:sSc.RBH.rp}}{=} \hspace{0.5mm} \frac{3l}{r_\sg} + \frac{6l^2}{r_\sg^2} + \mathcal{O}\big(l^3\big) \geqslant 0 .
\end{align}
As in the previously considered models, this expression cannot be zero unless $l=0$ and thus $r_-=0$ [cf.\ Eq.~\eqref{eq:sSc.RBH.rm}].

\subsection{Charged Hayward--Frolov model} \label{subsec:HF.RBH}
In Ref.~\citenum{f:16}, Frolov proposed a generalization of the Hayward model to include an electric charge $q$. In the generalized dynamical case, the metric function for this type of RBH is given by
\begin{align}
	f(v,r) &= 1 - \frac{\big(r_\sg(v) r - q(v)^2\big)r^2}{r^4+\big(r_\sg(v)r+q(v)^2\big)l(v)^2} ,
	\label{app:eq:ch_Hayw_Frol_RBH}
\end{align}
where, as in Subsec.~\ref{subsec:H.RBH}, $r_\sg$ and $l$ denote the Schwarzschild radius and minimal length scale, and the case $q=0$ reduces to the metric function of the uncharged dynamical Hayward model [Eq.~\eqref{eq:H.RBH.f(v,r)}]. The static case with $l=0$ reproduces the Reissner--Nordström metric. For both $q=l=0$, Eq.~\eqref{app:eq:ch_Hayw_Frol_RBH} corresponds to the Vaidya (or, in the static case, Schwarzschild) metric. 

As alluded to at the end of Sec.~\ref{sec:comp.cond}, the Reissner--Nordström metric belongs to the $k=1$ class of black hole solutions, and thus we must not use Eq.~\eqref{eq:KodamaSGv2} \textit{a priori}. The surface gravity of a Reissner-Nordström black hole is given by
\begin{align}
	\kappa_{\text{RN}} = \frac{r_+ - r_-}{2 r_+^2} , \label{eq:RNbh.SG}
\end{align}
where 
\begin{align}
	r_- &= m - \sqrt{m^2 - q^2} , \\
	r_+ &= m + \sqrt{m^2 - q^2} .
\end{align}
The dynamical generalization of the Reissner--Nordström metric is described by the metric function
\begin{align}
	f(v,r) = \frac{1}{r^2} \big(r-r_-(v)\big)\big(r-r_+(v)\big) ,
\end{align}
with the evolving inner and outer horizon specified by
\begin{align}
	r_-(v) &= m(v) - \sqrt{m(v)^2 - q(v)^2} , \label{eq:RNBH.dyn.rm} \\
	r_+(v) &= m(v) + \sqrt{m(v)^2 - q(v)^2} . \label{eq:RNBH.dyn.rp} 
\end{align}
Since this is a $k=0$ solution, we can make use of Eq.~\eqref{eq:KodSG} to determine the Kodama surface gravity at the outer horizon of an evolving Reissner--Nordström black hole and find
\begin{align}
	\kappa_{K_\text{RN}} = \frac{1}{2} \partial_r f(v,r) = \frac{r_+(v)-r_-(v)}{2 r_+(v)^2} .
	\label{app:eq:RN_KodSG}
\end{align}
On the other hand, from the charged Hayward--Frolov metric [Eq.~\eqref{app:eq:ch_Hayw_Frol_RBH}], we obtain
\begin{align}
	\kappa_{K_\text{HF}} = \frac{1-w_1(v,l)}{2r_+(v,l)} ,
	\label{app:eq:chHaywFrol_KodSG}
\end{align}
and the inner and outer horizon are given by
\begin{align}
	r_-(v,l) &= r_-(v) + \beta_-(v) l^2 + \mathcal{O}\big(l^3\big) , \label{eq:HF.RBH.rm(v,l)} \\
	r_+(v,l) &= r_+(v) + \beta_+(v) l^2+ \mathcal{O}\big(l^4\big) . \label{eq:HF.RBH.rp(v,l)}
\end{align}
Following the methodology of Sec.~\ref{sec:comp.cond}, considering the MS mass expansion allows us to identify the linear coefficient
\begin{align}
	w_1(v,l) &= \frac{q(v)^2}{r_+(v)^2} + \beta(v) l^2 + \mathcal{O}\big(l^4\big) , \label{app:eq:w1(v,l)}
\end{align}
where $\beta(v)$ denotes a lengthy coefficient that simplifies to $\beta(v) \to 3/r_\sg(v)^2$ in the uncharged case $q(v) \to 0$, in agreement with the expression derived for the uncharged dynamical Hayward model [cf.\ Eq.~\eqref{eq:H.RBH.w1}]. Substituting this result into Eq.~\eqref{app:eq:chHaywFrol_KodSG} and using Eqs.~\eqref{eq:HF.RBH.rm(v,l)}--\eqref{eq:HF.RBH.rp(v,l)}, we find
\begin{align}
	\kappa_{K_\text{HF}} = \frac{r_+(v)-r_-(v)}{2 r_+(v)^2} + \mathcal{O}\big(l^2\big) .
	\label{eq:HF.RBH.KodSG}
\end{align}
As can be seen by comparison with Eq.~\eqref{app:eq:RN_KodSG}, this expression matches that of the evolving Reissner--Nordström black hole only if $l=0$, in which case the horizons of the charged dynamical Hayward--Frolov RBH specified by Eqs.~\eqref{eq:HF.RBH.rm(v,l)}--\eqref{eq:HF.RBH.rp(v,l)} reduce to those of Eqs.~\eqref{eq:RNBH.dyn.rm}--\eqref{eq:RNBH.dyn.rp}. However, unlike the previously considered examples, the inner horizon $r_- \neq 0$ even if $l=0$ due to the presence of a charge term that is independent of $l$, cf.\ Eqs.~\eqref{eq:RNBH.dyn.rm} and \eqref{eq:HF.RBH.rm(v,l)}.

As evident from Eqs.~\eqref{app:eq:w1(v,l)} and \eqref{eq:HF.RBH.KodSG}, the condition for the compatibility of a dynamically evolving charged RBH with the first law of black hole mechanics is no longer encoded by the relation $w_1=0$. However, in the special circumstance where $l \to 0$, the new compatibility condition can be stated as
\begin{align}
	w_1(v,0) &= \frac{q(v)^2}{r_+(v)^2} .
\end{align}

\section{Testing regular black holes part \RNum{2}: inner-extremal model with a degenerate inner horizon} \label{sec:test.deg}
The inner-extremal RBH model proposed in Ref.~\citenum{cdlpv:22} solves the mass inflation instability problem at the expense of a degenerate inner horizon with vanishing surface gravity. As argued above, we consider its dynamical generalization. The metric function is given by choosing $a=3$, $b=1$ in Eq.~\eqref{eq:fvr.gen} such that in the generalized dynamical case
\begin{align}
	f(v,r) = g(v,r)\big(r-r_{-}(v)\big)^3\big(r-r_{+}(v)\big) , \label{eq:ieRBH.f(v,r)}
\end{align}
and by virtue of Eq.~\eqref{eq:f(v,r).poly.decomp} the choice $\sum_z \lambda_z r^z = 1 \Rightarrow m=0$ determines the degree $n-0=4=a+1$ of the polynomial in the denominator of Eq.~\eqref{eq:g(v,r).decomp}, i.e.\
\begin{align}
	g(v,r) &= \frac{1}{c_0 + c_1 r + c_2 r^2 + c_3 r^3 + c_4 r^4} , \label{eq:ieRBH.g(v,r)}
\end{align}
where we have once again omitted the dependencies of the coefficients $c_i \equiv c_i(r_-(v),r_+(v))$ for the sake of readability and will omit dependencies on $v$ in what follows as well. While $a=3$ for the model considered in Ref.~\citenum{cdlpv:22}, our derivation in what follows is valid for arbitrary RBH models with $m=0$ and $b=1$ [as otherwise the Kodama surface gravity would vanish at the outer horizon, cf.\ Eq.~\eqref{eq:KodSG.lim}], i.e.\ those where Eq.~\eqref{eq:g(v,r).decomp} admits the form
\begin{align}
	g(v,r) &= \frac{1}{c_0 + c_1 r + \cdots +c_{a+1} r^{a+1}} ,
	\label{eq:m0.g(v,r)}
\end{align}
where $c_{a+1}(r_-,r_+)= 1$ is required to recover the Vaidya form of the metric in the asymptotic limit $r \to \infty$. Based on dimensional grounds, the generic form of the coefficients $c_i(r_-,r_+)$ is prescribed by
\begin{align}
	c_i(r_-,r_+) = \sum\limits_{j=1}^\infty d_{ij} r_-^j r_+^{-j-i+(a+1)} \; \; \forall \; i \neq a+1 \; ,
	\label{eq:m0.coeff.form}
\end{align}
where the coefficients $d_{ij}$ are dimensionless\footnote{Note that the coefficients of our simple nonsingular spacetime [Subsec~\ref{subsec:sRBH}, Eq.~\eqref{eq:sRBH.c0c1}] are consistent with the required coefficient form specified in Eq.~\eqref{eq:m0.coeff.form} only if $r_-=0$ and $r_+=0$, i.e.\ if the spacetime has no horizons, in agreement with the trivial form of the metric function in Eq.~\eqref{eq:sRBH.f(v,r).triv}.}. If the inner horizon is absent ($r_- \to 0$), we should also recover the Vaidya form $f= 1 - r_+/r$ as the black hole center is then no longer regular. In this case, Eq.~\eqref{eq:fvr.gen} is given by
\begin{align}
	f(v,r) \big\vert_{r_- \to 0} = \frac{r^{a+1} \left(1 - \frac{r_+}{r}\right)}{c_0(0,r_+) + c_1(0,r_+) r + \cdots + r^{a+1}} .
\end{align}
Therefore, to recover the Vaidya form of the metric, we must have
\begin{align}
	c_i(0,r_+) = 0 \; \; \forall \; i \neq a+1 \; \Rightarrow \; g(v,r) \big\vert_{r_- \to 0} = \frac{1}{r^{a+1}} .
\end{align}
Using Eqs.~\eqref{eq:w1} and \eqref{eq:m0.g(v,r)}, the compatibility condition Eq.~\eqref{eq:w1=0} can be rewritten as
\begin{align}
	& \; g(v,r_+) r_+ \left( r_+ - r_- \right)^a = 1 , \\
	\Leftrightarrow \; & \left( r_+ - r_- \right)^a r_+ = \frac{1}{g(v,r_+)} \stackrel{\eqref{eq:m0.g(v,r)}}{=} \sum\limits_{i=0}^{a+1} c_i(r_-,r_+) r_+^i . \label{eq:m0.comp.cond}
\end{align}
Using the binomial theorem to expand the LHS, and Eq.~\eqref{eq:m0.coeff.form} for the coefficients on the RHS, Eq.~\eqref{eq:m0.comp.cond} can be rewritten as
\begin{equation}
\resizebox{0.99\linewidth}{!}{$
	\sum\limits_{k=0}^a 
	\begin{pmatrix}
		a \\ k
	\end{pmatrix}
	(-1)^{a-k} r_-^{a-k} r_+^{k+1} 
	=
	\sum\limits_{i=0}^{a} 
	\sum\limits_{j=1}^{\infty} 
	d_{ij} r_-^j r_+^{-j+(a+1)} + r_+^{a+1} .
$}
\end{equation}
Separating the $k=a$ term (whose contribution is $r_+^{a+1}$) from the summation on the LHS, this simplifies further to
\begin{align}
	\sum\limits_{k=0}^{a-1}
	\begin{pmatrix}
		a \\ k
	\end{pmatrix}
	(-r_-)^{a-k} r_+^{k+1} 
	=
	\sum\limits_{i=0}^a
	\sum\limits_{j=1}^{\infty} 
	d_{ij} r_-^j r_+^{-j+(a+1)} .
	\label{eq:m0.comp.cond.simp}
\end{align}
Note that the maximum possible exponent of $r_-$ on the LHS is $a$ (for $k=0$). Consequently, if Eq.~\eqref{eq:m0.comp.cond.simp} is treated as a polynomial with respect to $r_-$, the only way to have a solution is to truncate the summation over $j$ on the RHS at $j=a$. Furthermore, since $0 \leqslant k \leqslant a-1$ and $1 \leqslant j \leqslant a$, we can redefine the summation with respect to $k$ on the LHS and use the transformation $j \mapsto a-k$ to rewrite Eq.~\eqref{eq:m0.comp.cond.simp} as
\begin{align}
	\sum\limits_{j=1}^a
	\begin{pmatrix}
		a \\ a-j
	\end{pmatrix}
	(- r_-)^j r_+^{-j+(a+1)}
	=
	\sum\limits_{i=0}^a 
	\sum\limits_{j=1}^a 
	d_{ij} r_-^j r_+^{-j+(a+1)} .
\end{align}
As a result, the compatibility condition Eq.~\eqref{eq:w1=0} is satisfied if and only if
\begin{align}
	\sum\limits_{j=1}^a r_-^j \left[ 
	\begin{pmatrix}
		a \\ a-j
	\end{pmatrix}
	(-1)^j r_+^{-j+(a+1)} - \sum\limits_{i=0}^a d_{ij} r_+^{-j+(a+1)} 
	\right]
	= 0 .
	\label{eq:m0.comp.cond.brac}
\end{align}
We note that, in consonance with our analysis of nondegenerate RBH models in Sec.~\ref{sec:test.nondeg}, this relation is trivially satisfied if there is no inner horizon ($r_-\to0$). On the other hand, if an inner horizon is present ($r_-\neq0$) satisfying Eq.~\eqref{eq:m0.comp.cond.brac} requires
\begin{align}
	\boxed{
		\sum\limits_{i=0}^a d_{ij} 
		= 
		\begin{pmatrix}
			a \\ a-j
		\end{pmatrix}
		(-1)^j
		}
	\label{eq:m0.comp.cond.fin}
\end{align} 
This alternative form of the compatibility condition derived in Sec.~\ref{sec:comp.cond} allows us to test RBH models for which evaluating the expression given in Eq.~\eqref{eq:w1} is not straightforward due to the fact that $a>1$ and the coefficients in the polynomial decomposition of $g$ [Eq.~\eqref{eq:g(v,r).decomp}] are \textit{a priori} undetermined. With the commonly used assumptions of a positive MS mass $C>0$ [cf.\ Eq.~\eqref{eq:MSmass}] and regularity (expressed mathematically through the finiteness of relevant curvature scalars, such as the Ricci and Kretschmann scalar), it is possible to determine the lowest-order coefficients of Eq.~\eqref{eq:m0.g(v,r)}. However, this alone does not suffice to determine whether or not relation Eq.~\eqref{eq:m0.comp.cond.fin} is satisfied in general, as this would require the ability to determine every individual coefficient $c_i(r_-,r_+)$. Nonetheless, it is often possible to evaluate Eq.~\eqref{eq:m0.comp.cond.fin} in physically motivated scenarios where the degeneracy of the inner horizon $a$ in Eq.~\eqref{eq:fvr.gen} is known explicitly. 

We now examine the dynamical generalization [Eqs.~\eqref{eq:ieRBH.f(v,r)}--\eqref{eq:ieRBH.g(v,r)}] of the inner-extremal RBH model with $a=3$ proposed in Ref.~\citenum{cdlpv:22} and demonstrate that it cannot satisfy the condition Eq.~\eqref{eq:m0.comp.cond.fin} required to be compatible with the first law. In this model, $c_4=1$ and $c_3=-3r_-$ [cf.\ Eq.~\eqref{eq:ieRBH.g(v,r)}] are necessary to recover the Vaidya form of the metric asymptotically. The latter can be seen by expanding Eq.~\eqref{eq:ieRBH.f(v,r)} about the point $z \defeq 1/r = 0$ to represent the limit $r\to\infty$, which results in
\begin{align}
	f(v,r) \big\vert_{\tfrac{1}{r}=0} &= \frac{1}{c_4} - \frac{c_3 + c_4\big(3 r_- + r_+\big)}{c_4^2} \frac{1}{r} + \mathcal{O}\bigg(\hspace{-.25mm}\frac{1}{r^2}\hspace{-.5mm}\bigg) .
	\label{eq:ieRBH.f(v,r).lim}
\end{align}
Substitution of $c_4=1$ in Eq.~\eqref{eq:ieRBH.f(v,r).lim} and subsequent comparison with the Vaidya form $f=1-r_+/r$ yields $c_3=-3r_-$.
From the requirement of regularity, we obtain
\begin{align}
	c_0 = r_-^3 r_+ , \; \; c_1 = -r_-^2 (r_- + 3 r_+) ,
\end{align}
as can be verified, for instance, by evaluating the Ricci or the Kretschmann scalar.

The remaining coefficient $c_2$ can be determined from the requirement that $f$ be nondivergent, which in turn requires that its denominator 
\begin{align}
	D(v,r) & \defeq 1/g(v,r) \label{eq:ieRBH.D(v,r)} \\
	&\stackrel{\eqref{eq:ieRBH.g(v,r)}}{=} r_-^3 r_+ - r_-^2(r_- + 3r_+)r + c_2 r^2 -3r_- r^3 +  r^4 \nonumber
\end{align}
be nonzero. We rewrite Eq.~\eqref{eq:ieRBH.D(v,r)} as
\begin{align}
	\tilde{D}(v,r) &= \tilde{c}_2 r^2 + \Big(r - \frac{3 r_-}{2}\Big)^2 + r_-^3 r_+ \Big[1 - \frac{r}{2} \Big(\frac{3}{r_-}+\frac{1}{r_+}\Big)\Big]^2 ,
\end{align}
where
\begin{align}
	c_2 &= \tilde{c}_2 + \frac{15}{4} r_-^2 + \frac{9}{4} r_- r_+ + \frac{r_-^3}{4 r_+} , \; \; \tilde{c}_2 \geqslant 0 . \label{app:eq:c2_coeff}
\end{align}
Using Eq.~\eqref{eq:f.def.MSmass} to express the MS mass $C$ in terms of $f$ and expanding about the center $r=0$, we find
\begin{align}
	C(v,r) &= \frac{r_-^3 + 4 \tilde{c}_2 r_+ + 3 r_-^2 r_+ - 3 r_- r_+^2}{4 r_-^3 r_+^2} r^3 + \mathcal{O}(r^4) .
\end{align}
The positivity requirement for the MS mass then constrains the coefficient $\tilde{c}_2$ via
\begin{align}
	\begin{aligned}
		& r_-^3 + 4 \tilde{c}_2 r_+ + 3 r_-^2 r_+ - 3 r_- r_+^2 > 0 \\ 
		&\qquad \Rightarrow \; \tilde{c}_2 = \frac{\zeta - r_-^3 - 3 r_-^2 r_+ + 3 r_- r_+^2}{4 r_+} , \; \; \zeta > 0 . 
	\end{aligned}
	\label{app:eq:c_coeff}
\end{align}
Substituting Eq.~\eqref{app:eq:c_coeff} into Eq.~\eqref{app:eq:c2_coeff} yields
\begin{align}
	c_2 = \frac{\zeta}{4 r_+} + 3 r_-^2 + 3 r_- r_+ .
	\label{app:eq:c2_coeff_zeta}
\end{align}
This is all we need to explicitly identify the dimensionless coefficients $d_{ij}$ introduced in Eq.~\eqref{eq:m0.coeff.form}. We find
\begin{align}
	& c_0 = r_-^3 r_+ = d_{01} r_- r_+^3 + d_{02} r_-^2 r_+^2 + d_{03} r_-^3 r_+ \nonumber \\
	& \qquad \Rightarrow \; d_{01} = 0 \; , \; \; d_{02} = 0 \; , \; \; d_{03} = 1 , \label{app:eq:d0j} \\
	& c_1 = - r_-^3 - 3 r_-^2 r_+ = d_{11} r_- r_+^2 + d_{12} r_-^2 r_+ + d_{13} r_-^3 \nonumber \\
	& \qquad \Rightarrow \; d_{11} = 0 \; , \; \; d_{12} = -3 \; , \; \; d_{13} = - 1 , \label{app:eq:d1j} \\
	& c_3= - 3 r_- = d_{31} r_- + d_{32} r_-^2 r_+^{-1} + d_{33} r_-^3 r_+^{-2} \nonumber \\
	& \qquad \Rightarrow \; d_{31} = -3 \; , \; \; d_{32} = 0 \; , \; \; d_{33} = 0 . \label{app:eq:d3j} 
\end{align}
To proceed with the identification of the coefficients $d_{2j}$, we note that the dimensions of $\zeta$ are $[\zeta]=L^3$ (i.e.\ $\zeta$ is cubic in length), as can be seen from Eq.~\eqref{app:eq:c2_coeff_zeta}. Therefore, we can write the first term in Eq.~\eqref{app:eq:c2_coeff_zeta} as
\begin{align}
	\frac{\zeta}{4 r_+} &= \zeta_1 r_- r_+ + \zeta_2 r_-^2 + \zeta_3 r_-^3 r_+^{-1} > 0 ,
	\label{app:eq:zeta_i_pos}
\end{align}
where the coefficients $\zeta_i$ are dimensionless. Hence
\begin{align}
	\begin{aligned}
		c_2 &= d_{21} r_- r_+ + d_{22} r_-^2 + d_{23} r_-^3 r_+^{-1} \\
		&= (\zeta_1+3) r_- r_+ + (\zeta_2+3) r_-^2 + \zeta_3 r_-^3 r_+^{-1} \\
		& \quad \Rightarrow \; d_{21} = \zeta_1+3 \; , \; \; d_{22} = \zeta_2+3 \; , \; \; d_{23} = \zeta_3 .
	\end{aligned}
\end{align}
Next, we use Eq.~\eqref{eq:m0.comp.cond.fin} to determine the coefficients $\zeta_i$. We find
\begin{align}
	\sum\limits_{i=0}^3 d_{i1} &= 
	\begin{pmatrix}
		3 \\ 2
	\end{pmatrix}
	(-1)^1 = -3 \nonumber \\
	&= d_{01} + d_{11} + d_{21} + d_{31} \; \Rightarrow \; \zeta_1 = -3 , 
	\label{app:eq:zeta1} \\
	\sum\limits_{i=0}^3 d_{i2} &= 
	\begin{pmatrix}
		3 \\ 1
	\end{pmatrix}
	(-1)^2 = 3 \nonumber \\
	&= d_{02} + d_{12} + d_{22} + d_{32} \; \Rightarrow \; \zeta_2 = 3 , 
	\label{app:eq:zeta2} \\
	\sum\limits_{i=0}^3 d_{i3} &= 
	\begin{pmatrix}
		3 \\ 0
	\end{pmatrix}
	(-1)^3 = -1 \nonumber \\
	&= d_{03} + d_{13} + d_{23} + d_{33} \; \Rightarrow \; \zeta_3 = -1 .
	\label{app:eq:zeta3}
\end{align}
Substituting these values into Eq.~\eqref{app:eq:zeta_i_pos}, we obtain
\begin{align}
	\begin{aligned}
		-3 r_- r_+ + 3 r_-^2 - r_-^3 r_+^{-1} > 0 \\ 
		\qquad \Leftrightarrow \; 3 r_+^2 - 3 r_- r_+ + r_-^2 < 0 .
	\end{aligned}
	\label{app:eq:rp_poly_ineq}
\end{align}
Considering Eq.~\eqref{app:eq:rp_poly_ineq} as a polynomial in $r_+$, its discriminant is given by $-3r_-^2 < 0$, which implies that the polynomial has two distinct complex conjugate roots and is thus always positive. However, this is in direct contradiction with the inequality in the second line of Eq.~\eqref{app:eq:rp_poly_ineq}, indicating that it is impossible for the inner-extremal RBH model to satisfy the necessary relation Eq.~\eqref{eq:m0.comp.cond.fin} that is required to be compatible with the first law of black hole mechanics.

\section{Page evaporation law} \label{sec:Page.law}
Building on Hawking's result \cite{h:75}, Page demonstrated that the mass loss due to the emission of Hawking radiation is described by the formula
\begin{align}
	\frac{dM}{dt} = - \hspace{-1.5mm} \sum\limits_{j,\ell,m,p} \frac{1}{2\pi} \int\limits_0^\infty \frac{\omega \Gamma_{j \omega \ell m p}}{e^{2 \pi \omega/\kappa} - 1} d\omega ,
	\label{eq:Page_formula}
\end{align}
where $j$ labels the emitted particle species, and $\Gamma_{j \omega \ell m p}$ denotes the absorption probability\footnote{In practice, these are determined using analytical and numerical techniques in a formalism originally developed by Teukolsky and Press \cite{t:73,tp:74}.} for an incoming wave mode labeled by the frequency $\omega$, spheroidal harmonic $\ell$, axial quantum number $m$, and polarization $p$. Although this formula accounts for both angular momentum and charge, an isolated black hole is expected to be well approximated by an uncharged spherically symmetric metric due to the fact that in the Hawking process angular momentum is shed much faster than mass \cite{p:76}, and charged black holes rapidly discharge in a Schwinger-like pair production process \cite{s:51,g:75}.

Since the ingoing Vaidya metric with decreasing mass ($C'(v)<0$) is the appropriate choice to model the effects of Hawking radiation (see Table 2 in Ref.~\citenum{mmt:rev:22}), Eq.~\eqref{eq:Page_formula} is often expressed in advanced null coordinates $(v,r)$ using several physically motivated simplifying assumptions (most notably the restriction to $\ell=m=0$ modes and the approximate relation $\Gamma \simeq \omega^2 r_\sg^2$; see Refs.~\citenum{w:94,pt:book:09,h:16} for a more detailed account) as
\begin{align}
	\frac{dM}{dv} \simeq - \frac{a}{M^2} \, \, \Leftrightarrow \, \, \frac{dr_+}{dv} \simeq - \frac{\alpha}{r_+^2} \, ,
\end{align}
which implies that a black hole of initial mass $M_0$ will evaporate in a time $t_e \sim M_0^3$, and the explicit form of the coefficients and their expansion about $w_1=0$ are given by
\begin{align}
	\alpha = 8a = - \frac{4}{\pi} \frac{1}{e^{\tfrac{4\pi}{1-w_1}} - 1} , \\
	\alpha = - \frac{4}{\pi} \frac{1}{e^{4\pi} - 1} + \mathcal{O}(w_1) ,
\end{align}
respectively. Consequently, the standard Page evaporation law is modified if the condition $w_1=0$ derived in Sec.~\ref{sec:comp.cond} is not satisfied.

\section{Conclusions} \label{sec:concl}
Based on the assertion that the surface gravity of an evolving black hole horizon should approach the expression prescribed by the first law of black hole mechanics [Eq.~\eqref{eq:firstlaw_sg}] in the quasistatic limit, we have derived a compatibility condition for generic spherically symmetric dynamical black holes [Sec.~\ref{sec:comp.cond}, Eq.~\eqref{eq:w1=0}]. In our analysis of the dynamical generalizations of RBH models typically considered in the literature, we have evaluated the compatibility condition explicitly for the respective metric functions that describe them, and demonstrated that none of them satisfies the necessary condition required to be compatible with the conventional form of the first law of black hole mechanics (Sec.~\ref{sec:test.nondeg} and Sec.~\ref{sec:test.deg}). As outlined in Sec.~\ref{sec:Page.law}, this also implies that --- if the decrease in mass $\delta M < 0$ due to the emission of Hawking radiation is indeed proportional to the surface gravity as stipulated by the first law --- then the dynamical evolution of such RBHs cannot be accurately described by the standard Page evaporation law. One may argue that this is a somewhat counterintuitive result, considering that the derivation of Eq.~\eqref{eq:Page_formula} is based on Hawking fluxes perceived in the asymptotic limit, and thus one would naively expect that the minimal length scale introduced for the purpose of regularization should not affect the outcome.

Our analysis suggests that the incompatibility of dynamical RBHs with the first law of black hole mechanics is directly linked to the minimal length scale $l$ (which can be interpreted as an additional hair) introduced for the regularization and the presence of an inner horizon, which are the main characteristics (together with their regular center) that distinguish RBH models from alternative descriptions of trapped spacetime domains. Since both are necessary ingredients in the regularization procedure to avoid singularities, one may conjecture that there is a more fundamental physical or topological principle at play that prevents nonsingular black hole spacetimes from satisfying the first law. 

In their most conservative form, the conclusions of our analysis may be stated as follows: nonsingular black holes are incompatible with the widely accepted semiclassical description of evaporating black holes that is based on the results of Refs.~\citenum{bch:73} and \citenum{h:74,h:75,p:76}. Unless one is willing to give up either the idea of regularity and an interior that is physically well behaved all the way down to the center or the first law of black hole mechanics (and its associated thermodynamic interpretation of surface gravity as an effective temperature), our results demonstrate that modifications of the first law are required even at the level of semiclassical gravity.

We note that our analysis is consistent with the interpretation of the deviation from the standard form of the first law [Eq.~\eqref{eq:first_law}] as a thermodynamic pressure term as has been proposed, for instance, in Refs.~\citenum{h:98} and \citenum{lm:22}, which can be seen from Eqs.~\eqref{eq:first_law_extended}--\eqref{eq:pressure_term}. In this sense, the linear coefficient of the MS mass $w_1$ encodes rather specific information about the thermodynamic properties of black holes. In fact, as evident from Eq.~\eqref{eq:first_law_extended}, knowledge of $w_1$ suffices to fully specify the generalized first law of black hole mechanics.

\acknowledgments
We would like to thank Yasha Neiman, Fil Simovic, and Daniel Terno for useful discussions and helpful comments. SM is supported by the Quantum Gravity Unit of the Okinawa Institute of Science and Technology (OIST). IS is supported by an International Macquarie University Research Excellence Scholarship (IMQRES).

\end{document}